\documentclass[pra,aps,twocolumn,showpacs,showkeys,floatfix,superscriptaddress]{revtex4-1}
\usepackage{color}
\usepackage{graphicx,amssymb,epsfig,epsf,amsmath} 
\usepackage[utf8]{inputenc}

\begin{document}
\title{Sorting Fermionization from Crystallization in Many-Boson Wavefunctions}
 \author{S. Bera}
  
\affiliation{Department of Physics, Presidency University, 86/1 College Street, Kolkata 700 073, India.}
\author{B. Chakrabarti}
\affiliation{Instituto de F\'isica, Universidade de S\~{a}o Paulo, CEP 05508-090, S\~{a}o Paulo, Brazil }
\affiliation{Department of Physics, Presidency University, 86/1 College Street, Kolkata 700 073, India.}

\author{A. Gammal}
\affiliation{Instituto de F\'isica, Universidade de S\~{a}o Paulo, CEP 05508-090, S\~{a}o Paulo, Brazil }

\author{M. C. Tsatsos}
\affiliation{Instituto de F\'sica de S\~ao Carlos, Universidade de S\~ao Paulo,
CP 369,13560-970, S\~ao Carlos, SP, Brasil}

\author{M. L. Lekala}
\affiliation{Department of Physics, University of South Africa P.O. Box-392, Pretoria 0003, South Africa}

\author{B. Chatterjee}
\affiliation{Department of Physics, Indian Institute of Technology-Kanpur, Kanpur 208016, India}

\author{C. L\'ev\^eque}
\affiliation{Wolfgang Pauli Institute c/o Faculty of Mathematics, University of Vienna, Oskar-Morgenstern Platz 1, 1090 Vienna, Austria}
\affiliation{Vienna Center for Quantum Science and Technology, Atominstitut, TU Wien, Stadionallee 2, 1020 Vienna, Austria}

\author{A. U. J. Lode}
\affiliation{Wolfgang Pauli Institute c/o Faculty of Mathematics, University of Vienna, Oskar-Morgenstern Platz 1, 1090 Vienna, Austria}
\affiliation{Vienna Center for Quantum Science and Technology, Atominstitut, TU Wien, Stadionallee 2, 1020 Vienna, Austria}
\affiliation{Institute of Physics, Albert-Ludwig University of Freiburg, Hermann-Herder-Strasse 3, 79104 Freiburg, Germany}

\date{\today}

\begin{abstract} Fermionization is what happens to the state of strongly interacting repulsive bosons interacting with contact interactions in one spatial dimension. Crystallization is what happens for sufficiently strongly interacting repulsive bosons with dipolar interactions in one spatial dimension. Crystallization and fermionization resemble each other: in both cases -- due to their repulsion -- the bosons try to minimize their spatial overlap. We trace these two hallmark phases of strongly correlated one-dimensional bosonic systems by exploring their ground state properties using the one- and two-body density matrix. We solve the $N$-body Schr\"odinger equation accurately and from first principles using the multiconfigurational time-dependent Hartree for bosons (MCTDHB) and for fermions (MCTDHF) methods. Using the one- and two-body density, fermionization can be distinguished from crystallization in position space. For $N$ interacting bosons, a splitting into an $N$-fold pattern in the one-body and two-body density is a unique feature of both, fermionization and crystallization. We demonstrate that the splitting is incomplete for fermionized bosons and restricted by the confinement potential. This incomplete splitting is a consequence of the convergence of the energy in the limit of infinite repulsion and is in agreement with complementary results that we obtain for fermions using MCTDHF.  For crystalline bosons, in contrast, the splitting is complete: the interaction energy is capable of overcoming the confinement potential. Our results suggest that the spreading of the density as a function of the dipolar interaction strength diverges as a power law. We describe how to distinguish fermionization from crystallization experimentally from measurements 
of the one- and two-body density.
\end{abstract}
\maketitle
\section{Introduction}

The physics of the ultracold Bose gas in one spatial dimension is strongly different from that of its three-dimensional  counterpart~\cite{Petrov:00,Dunjko:01}. In one spatial dimension, quantum effects are much more prominent. When the interactions are strong, quantum fluctuation are enhanced. Experimentally, in cold atom systems, the dimensionality can be manipulated using a tight transversal confinement that essentially 
freezes the radial degrees of freedom~\cite{Gorlitz:01,Greiner:01}. Such quasi-one dimensional systems display intriguing physics that cannot be realized for three-dimensional systems: \textit{Fermionization}, occurs for strongly interacting bosons with contact interactions \cite{Girardeau:60,Zurn:12,Jacqmin:11,Alon:05,Zollner:06,Zollner:08,Roy:18,Kocik:17b} and \textit{crystallization} 
emerges for sufficiently strongly interacting bosons with dipole-dipole interactions~\cite{Zollner:11,Zollner:11pra,astrakharchik:08a,astrakharchik:08b,Deuretzbacher:10,arkhipov05,Sowinski:19}.
%888
For bosons with contact interactions, fermionization leads to the formation of the Tonks-Girardeau (TG) gas
when the interaction strength tends to infinity. This is a consequence of the Bose-Fermi 
mapping~\cite{Girardeau:60,Paredes:04,Kinoshita:04,Deuretzbacher:07} which implies that strongly interacting bosons and non-interacting spinless fermions have the same one-body density in position space. With increasing interaction strength, not only the density, but also 
the energy of fermionized bosons saturates to the energy of non-interacting fermions.

In the case of dipolar interactions, the remarkable phenomenon of crystallization occurs when the interaction strength is moderately large. Bosons interacting via a dipole-dipole interaction potential have 
become the primary cold atom system to investigate the many-body physics triggered as consequence of long-range interactions~\cite{Santos:03,Andre:06,Buchler:07,baranov08,griesmaier05,beaufils08}. The long-ranged and anisotropic nature of the dipolar 
interaction potential results in a variety of interesting effects and phenomena~\cite{Lahaye:09}, like 
crystallization in one- and two-dimensional systems~\cite{Zollner:11,Zollner:11pra,astrakharchik:08a,astrakharchik:08b,Deuretzbacher:10,arkhipov05,Romanovsky:04,Kocik:17,Kocik:15}, that are completely different from the emergent phenomena in the case of strong contact interactions. Crystallization is a consequence of the repulsive and 
long-ranged tail of the dipolar interactions dominating the physics \cite{Chatterjee:12,Chatterjee:17}: the bosons maximally separate and minimize their spatial overlap. Unlike in the fermionization of bosons with strong contact interactions, the energy of crystallized bosons does not saturate. We note here that it is formally possible to define and measure an order parameter which is a function of the eigenvalues of the reduced one-body density matrix that allows to unequivocally identify the crystal phase of one-dimensional dipolar bosons~\cite{Chatterjee:17}. Furthermore, we note that the formation of a crystal state is a generic feature of many-body systems of particles with long-ranged interactions. Fermions with long-ranged interactions, for instance, form a so-called Wigner crystal~\cite{Wigner,Sowinski:19}.

In this work, we analyze the differences between fermionized and crystallized bosons' wavefunctions using the energy as well as the
one-body and two-body reduced density matrix. We demonstrate how fermionization can be distinguished from crystallization by quantifying the (experimentally accessible) spread of the one-body and two-body densities. The different spreading characteristics of the one-body and two-body densities for fermionized as compared to crystallized bosons are a direct consequence of the different behavior of the energy as a function of interaction strengths. For dipolar interactions the energy as a function of interaction strength is unbounded; this is in stark contrast to the bounded energy as a function of interactions for contact interactions. Current experimental setups, for instance, for Erbium~\cite{Ferlaino:12}, Erbium-Erbium molecules~\cite{Ferlaino:15}, or Sodium-Potassium molecules~\cite{Zwierlein:15}, enable the experimental exploration of systems with dominant dipole-dipole interactions needed for probing the physics of crystallization. 

The (momentum) densities of bosons with dipolar interactions have been compared to those of fermions with dipolar interactions in Ref.~\cite{Deuretzbacher:10}. Going beyond Ref.~\cite{Deuretzbacher:10}, we compare  and quantify the spreading of the full density matrices of bosons with dipolar interactions to the spreading of the density matrices of bosons with contact interactions.
We note that Ref.~\cite{CONDMAT} discusses and compares the physics of spin-1/2 fermions with contact and with long-ranged interactions in lattices using a Hubbard description. Our work complements the findings in Ref.~\cite{CONDMAT} by providing a comparison of single-component (``spin-0'') bosons with contact and long-ranged dipolar interactions. 

Fermionization and crystallization entail the breakdown of mean-field approaches like the time-dependent Gross-Pitaevskii (GP) equation \cite{Pethick:02,Bogoliubov:91,Pitaevskii:03}. To go beyond the GP approximation, multiconfigurational methods are employed \cite{MCTDHB1, MCTDHB2, MCTDHF, Cao:13, Leveque:17, Leveque:18, Lode:19}. Variational calculation using parametrized gaussian functions as single particle states have been successfully applied to investigate the crystallization of few particles in two dimensions ~\cite{Romanovsky:04,Yannouleas:07,Mese:11}. Here, we use the multiconfigurational time-dependent Hartree for bosons 
(MCTDHB)~\cite{MCTDHB1,MCTDHB2} and fermions (MCTDHF)~\cite{MCTDHF} methods implemented in the MCTDH-X software package~\cite{MCTDHF,Axel:Thesis,Axel:Spin,ultracold} to compute the ground state of the few-particle Schr\"odinger equation, see Ref.~\cite{Lode:19} for a Review. While the MCTDHB method aims at solving the time-dependent Schr\"odinger equation for a many-body system, using imaginary time-propagation provides the ground-state of the system variationally, equivalent to the work of Ref.~\cite{Heimsoth:10}.

We illustrate our findings with computations for $N=4$ bosons in a parabolic trapping potential and trace the complete range of dipolar and contact interaction strengths by obtaining highly accurate results with MCTDHB. 

This paper is structured as follows: in Sec.~\ref{sec:hamiltonian}, we discuss the Hamiltonian and quantities of interest, in Sec.~\ref{sec:method}, we introduce the numerical method, MCTDHB, that we use for obtaining solutions of the few-body Schrödinger equation, in Sec.~\ref{sec:results} we analyze fermionized and crystallized bosonic few-body states and discuss how they can be sorted from each other and we conclude our paper in Sec.~\ref{sec:conclusions}. Results for other observables and an assessment of the accuracy of our computations with the exact diagonalization and MCTDHF approaches are collected in the Appendices~\ref{MCTDHBvsED} and \ref{diffN}.

\section{Hamiltonian, one- and two-body density}\label{sec:hamiltonian}
 
In order to discuss the stationary properties of the ground state (GS) of crystalline and fermionized bosons, we consider the time-independent many-body Schrödinger equation,
\begin{equation}
 \hat{H} \vert \Psi \rangle \label{TISE} =  E|\Psi \rangle 
 \end{equation}
Here, $\vert \Psi \rangle$ is the many-body ground state, $E$ its energy, and $\hat{H}$ the $N$-particle Hamiltonian in dimensionless units~\cite{units},
 \begin{equation}
H= \sum_{i=1}^{N} \left( - \frac{1}{2} \frac{\partial^{2}}{\partial x_i^{2}} + V(x_i) \right) + \sum_{i<j}W(x_i-x_j),\label{HAM}
\end{equation}
where we set $V(x_i) = \frac{1}{2} x_i^{2}$ to be the external harmonic trap and $W(x_i-x_j)$ the interaction potential. All quantities are dimensionless and expressed in harmonic oscillator units. To ensure that the system is in the quasi-1D regime, we assume strong confinement frequencies in the transversal direction, providing a cigar-shaped atomic density. The contact interactions read, 
\begin{equation}
W(x_i-x_j) = \lambda \delta(x_i-x_j),
\end{equation}
where $\lambda$ is the interaction strength determined by the scattering length $a_s$ and the transverse confinement frequencies~\cite{Olshanii:98}. For long-ranged dipolar interactions, we have  
\begin{equation}
 W(x_i-x_j) = \frac{g_d} {|x_i-x_j|^{3} + \alpha}, \label{ddint}
\end{equation}
where $g_d$ is the dipolar interaction strength and $\alpha$ is a short-range cut-off to avoid the divergence at $x_i = x_j$. Repulsive interactions could be obtained in a quasi-1D BEC by imposing an external magnetic field to align all the dipole moments of the atoms ~\cite{Deuretzbacher:10}.    
This simple approximation to the one-dimensional dipole-dipole interaction potential in Eq.~\eqref{ddint} is justified for the moderate to large interaction strengths and large inter-particle distance with respect to the harmonic-length of the transversal confinement~\cite{Sinha:07,Deuretzbacher:10,Zollner:11,Zollner:11pra,Sowinski:19}, that we focus on in the present work. For such interaction strengths the dipole-dipole interaction potential is well-approximated by the $\vert x_i-x_j \vert^{-3}$ tail in Eq.~\eqref{ddint}, see ~\cite{cai10}. Moreover, we have verified the consistency of the approximation in Eq.~\eqref{ddint} for the same choice of cutoff parameter, $\alpha=0.05$, by a direct comparison to a dipole-dipole interaction augmented with an additional contact interaction potential, see Ref.~\cite{Chatterjee:17}. A rigorous discussion of the dipole-dipole interaction potential in one and two spatial dimensions can be found in Ref.~\cite{cai10}. Here, for the sake of computational complexity, we will focus on quasi-one-dimensional systems, with $N=4$ interacting bosons for all our calculations and consider repulsive interactions, i.e., $\lambda>0$ and $g_{d}>0$, exclusively  .

In the following we discuss the reduced one-body density matrix, defined as
\begin{equation}
\rho^{(1)}(x, x^{\prime})= \langle \Psi | {\hat\psi}^{\dagger}(x^{\prime}) {\hat\psi}(x) | \Psi \rangle. \label{1RDM}
\end{equation} 
Its diagonal, 
\begin{equation}
\rho(x) \equiv \rho^{(1)}(x,x^{\prime}=x) \label{1RDMdiag}
\end{equation}
is simply the one-body density. As a precursor of correlation effects  that may be present in the state $\vert \Psi \rangle$ of the system, we use the eigenvalues $\rho_i^{(NO)}$ of the reduced one-body density matrix $\rho^{(1)}$ in Eq.~\eqref{1RDM}. For this purpose, we write $\rho^{(1)}$ in its eigenbasis:
\begin{equation}
 \rho^{(1)}(x,x') = \sum_i \rho_i^{(NO)} \Phi^*_i(x') \Phi_i(x).\label{NOs}
\end{equation}
The eigenvalues $\rho_i^{(NO)}$ and eigenfunctions $\Phi_i(x)$ are referred to as natural occupations and natural orbitals, respectively.
If only a single eigenvalue $\rho_i^{(NO)}$ is macroscopic, then the state $\vert \Psi \rangle$ describes a Bose-Einstein condensate~\cite{Penrose:56}.
The case when multiple eigenvalues $\rho_i^{(NO)}$ are comparable to the number of particles $N$ is referred to as fragmentation~\cite{Axel:Spin,Axel:Thesis,Lode:17,MCTDHB1,MCTDHB2,Nozieres:82,Spekkens:99}.

In the following, we will also use the two-body density $\rho^{(2)}$ to characterize crystallization and fermionization.
It is defined as 
\begin{equation}
\rho^{(2)}(x_1,x_2) = \langle \Psi | {\hat\psi}^{\dagger}(x_1) {\hat\psi}^{\dagger}(x_2) {\hat\psi}(x_1) {\hat\psi}(x_2) | \Psi \rangle.\label{2RDM}
\end{equation}
The two-body density quantifies the probability to detect two particles at positions $x_1$ and $x_2$. 

\section{Numerical Method}\label{sec:method}
The computation of the exact many-body wave function is a difficult problem. To attack the many-body Schrödinger equation, Eq.~\eqref{TISE}, we use the time-dependent Schrödinger equation,
\begin{equation}
 i\partial_t \vert \Psi \rangle = \hat{H} \vert \Psi \rangle.\label{TDSE}
 \end{equation}
with a Wick's rotation $t\rightarrow -i\tau$, i.e., a propagation with imaginary time. We expand the many-body wavefunction $\vert \Psi \rangle$ of $N$ interacting bosons in a complete set of time-dependent permanents $\vert \vec{n};t\rangle=\vert n_1,...,n_M;t\rangle$ with at most $M$ single-particle states or orbitals. The MCTDHB ansatz for the many-body wave function is thus
\begin{equation}
\lvert \Psi(t)\rangle = \sum_{\vec{n}}^{} C_{\vec{n}}(t)\lvert \vec{n};t\rangle
\label{many_body_wf}.
\end{equation}
Here, the permanents $\lvert \vec{n};t\rangle$ are symmetrized bosonic many-body states that are also referred to as ``configurations''. The sum in Eq.~\eqref{many_body_wf} runs on all configurations $\vec{n}$ of $N$ particles in $M$ orbitals. The number of permanents and coefficients $C_{\vec{n}}(t)$ is
 $N_{conf}$= $\left(\begin{array}{c} N+M-1 \\ N \end{array}\right)$.
In the second quantized representation the permanents are given as
\begin{equation}
\lvert \bar{n};t\rangle = \lvert n_{1},...n_{M};t\rangle = \prod_{i=1}^{M}
\left( \frac{ \left( \hat{b}_{i}^{\dagger}(t) \right)^{n_{i}} } {\sqrt{n_{i}!}} \right) \lvert vac \rangle.
\label{2nd_quantized_wf}
\end{equation} 
Here $\hat{b}_{k}^{\dagger}(t)$ is the bosonic creation operator which creates a boson in the time-dependent single particle state $\phi_{k}(\vec{r},t)$. 
Eq.~\eqref{many_body_wf} spans the full $N$-body Hilbert space in the limit of $M\rightarrow \infty$. For practical computations, we restrict the number of orbitals and require the convergence of our observables, like the one- and two-body density matrix, with respect to the number of single-particle states $M$. 

A set of coupled equations of motion for, both, the time-dependent expansion coefficients $C_{\vec{n}}(t)$ and the time-dependent orbitals $\phi_{k}(\vec{r},t)$ are obtained by requiring the stationarity of the action of the time-dependent Schrödinger equation~\cite{MCTDHB1,MCTDHB2} under variations of $C_{\vec{n}}(t)$ and $\phi_{k}(\vec{r},t)$. Using MCTDHB, both, the coefficients and orbitals are variationally optimized~\cite{TDVP}. MCTDHB is thus fundamentally different from exact diagonalization, i.e., an ansatz built with \textit{time-independent} orbitals. It can be demonstrated that MCTDHB delivers solutions of the Schrödinger equation at a significantly increased accuracy in comparison to exact diagonalization approaches when the same number of single-particle basis states is employed, see Ref.~\cite{Axel:Thesis,Axel:HIM} for a demonstration with the harmonic interaction model and Appendix~\ref{MCTDHBvsED} for a demonstration with dipole-dipole interactions, i.e., the Hamiltonian in Eqs.~\eqref{HAM} and \eqref{ddint}. 
Despite the accuracy of MCTDHB for weakly interacting particles, for strong interactions a large number of orbitals is required to describe the system accurately. In the case $\lambda \rightarrow \infty$, the Bose-Fermi mapping provides analytical solutions that can be compared to the numerical results. These difficulties to converge MCTDHB results for strong contact interactions may be an instance of the discussion provided in Ref.~\cite{Gwak:18}. 

We solve the set of coupled MCTDHB equations using the MCTDH-X software~\cite{MCTDHF,Axel:Spin,ultracold,Axel:Thesis}. When imaginary time is used, the propagation of an initial guess function converges to the ground state of the system, and the stationary properties of the system can be investigated. 

\section{Fermionization \textit{vs} Crystallization}\label{sec:results}

We now discuss our findings on the fermionized and the crystalline state of parabolically trapped one-dimensional ultracold bosons. We first independently characterize fermionization and crystallization from a ``many-body point of view'', see Sec.~\ref{sec:fermionization} and Sec.~\ref{sec:crystallization},  respectively. Thereafter, we investigate how to sort the one, fermionization, from the other, crystallization, in Sec.~\ref{sec:CXF}. Here and in the following, we used the term ``many-body point of view'' to highlight that our considerations go beyond an effective single-particle or mean-field description of the state.

\subsection{Fermionization}\label{sec:fermionization}
Bosons fermionize when they feel an infinitely repulsive contact interaction in one spatial dimension. For fermionized bosons, the total energy $E$ and the density [Eq.~\eqref{1RDMdiag}] of the system become exactly equal to the energy and the density of non-interacting spinless fermions, respectively. For our showcase of few-bosons system $N=2$ to $N=5$ bosons in a harmonic trap with frequency one, $V(x)=\frac{1}{2}x^2$, the limiting value is thus $E_{\lambda\rightarrow\infty}^{N}=\frac{N^{2}}{2}$. 

We start our investigation with the one-body density as a function of interaction strength $\lambda$ [Fig.~\ref{Fig1}(a-b)]. For comparatively weak repulsion, the density is clustered at the center of the trap, but becomes flatter and broader when $\lambda$ increases. For stronger repulsion, the density gradually acquires modulations and the number of humps finally saturates to the number of bosons in the system; four humps for $N=4$ bosons are clearly visible when the interaction strength goes above $\lambda \sim 10$. The emergence of $N$ maxima in the density indicates that the TG regime is approached.  
The density modulations/humps are more pronounced in the center of the trap, where the potential is close to zero. For larger distances from the origin, the humps in the density are less pronounced due to the non-zero value of the confinement potential. 
Importantly, the outermost density modulation also becomes less pronounced if the number of particles $N$ is increased. See also Appendix~\ref{diffN} for a direct comparison of the relative height of innermost and outermost peaks for different particle numbers $N$.

We note that the density's maxima in the Tonks-Girardeau regime are distinct but {\it not isolated}. We also observe that, once the TG regime is reached, the density does not broaden further with increasing values of $\lambda$ for all particle numbers. We also provides a direct comparison with the ground state properties of non-interacting fermions computed with the multiconfigurational time-dependent Hartree method for fermions (MCTDHF), see Fig.~\ref{Fig5}. We note that the results for non-interacting fermions can be obtained analytically, i.e., here, we use the heavy MCTDHF method only for the sake of computational convenience. 

We now move to discuss the two-body densities $\rho^{(2)}$ of bosons with contact interactions [see Fig.~\ref{Fig2}(a)].
For weak interaction strength, $\lambda=0.1$, the bosons are clustered near the center, i.e,
$x_1=x_2=0$ [Fig.~\ref{Fig2}(a)]. As the interaction strength increases, $\rho^{(2)}$ spreads out to the off-diagonal $(x_1 \neq 
x_2)$ while the diagonal ($x_1\sim x_2$) is depleted [see Fig.~\ref{Fig2}(a) for $\lambda=1$].

For stronger repulsion a so-called ``correlation hole'' in the two-body density forms on the diagonal, $\rho^{(2)}(x,x) \rightarrow 0$ [see Fig.~\ref{Fig2}(a) for $\lambda=10$ and $\lambda=30$]. The probability of finding two bosons at the same position tends towards zero. In the limit of infinite repulsion the correlation hole persists in $\rho^{(2)}$. 
In analogy, however, to the boundedness of the energy as a function of the interaction strength, the width of two-body density on its anti-diagonal [$\rho^{(2)}(x,-x)$] is also bounded, i.e., the spread of $\rho^{(2)}$ converges in the fermionization limit when $\lambda\rightarrow\infty$. 

Similar to the one-body density, the maxima which are formed in the off-diagonal of the two-body density are distinct but not isolated [see Fig.~\ref{Fig2}(a) for $\lambda=10$ and $\lambda=30$]. 

We infer that {\it { the correlation hole along the diagonal}} and {\it {the confined spread}} are the unique signatures of the two-body density of a fermionized state.

\subsection{Crystallization}\label{sec:crystallization}

For bosons with dipole-dipole interactions, crystallization occurs when the long-range tail of the interaction [see Eq.~\eqref{ddint}] becomes dominant~\cite{Chatterjee:17}: the bosons form a lattice structure which allows them to minimize their mutual overlap. To characterize crystallization we analyze the one-body and two-body density for bosons with dipolar interaction of strength $g_d$. We choose the cut-off parameter $\alpha=0.05$ in Eq.~\eqref{ddint} such that the effective interaction such that
the effective interaction features the same physical beahavior as the ``real'' dipolar interaction that additionally contains a contact-interaction term (see Ref.~\cite{Chatterjee:17} for a direct comparison).

We plot the one-body density of $N=4$ bosons as a function of $g_d$ in Fig.~\ref{Fig1}(c) and (d). The system is condensed at the center of the trap for small $g_d$. As $g_d$ increases, the density starts to exhibit a four-hump structure (see Fig.~\ref{Fig1}(c) and (d) for $g_d \in [\sim 1, \sim 5]$) similar to the density observed for the fermionization of bosons with contact interactions [Fig.~\ref{Fig1}(a)]. 

This \textit{attempted fermionization} results from a dominant contribution of the short-range part of the dipolar interaction potential, see also Ref.~\cite{Deuretzbacher:10}. However, this fermionization-like behavior is only a precursor to the crystal transition that takes place when the long-range nature of the interaction starts to dominate the physics of the system for larger interaction strengths [Fig.~\ref{Fig1}(c) and (d) for $g_d\gtrsim10$]. For crystallized dipolar bosons at large $g_d$, the value of the density at its minima between the humps tends to zero while the spreading of the density profile diverges as $g_d$ increases, see Fig.~\ref{Fig8}. At $g_d=30.0$, we observe four \textit{well-isolated} peaks heralding the crystallization of the $N=4$ bosons. We collect results for other numbers of bosons ($N=2,3,5,6$) with dipole-dipole interactions -- including the relative height of the peaks in the density that shows that the peaks are well-isolated in comparison to particles with contact interactions -- in Appendix~\ref{diffN}. A comparison of MCTDHB results with exact diagonalization is shown in Appendix~\ref{MCTDHBvsED}.

We now analyze the two-body density for dipole-dipole interactions [Fig.~\ref{Fig2}(b)]. For small interaction strength, $g_d=0.1$, the atoms are clustered together at the center of the trap. As $g_d$ increases, a correlation hole develops: $\rho^{(2)}(x,x)$ tends to zero [Fig.~\ref{Fig2}(b) for $g_d\geq1$]. Thus, due to the long-range interaction, the probability of finding two bosons in the same place is strongly reduced. In the crystalline phase [Fig.~\ref{Fig2}(b) for $g_d\geq10$]: the bosons escape their spatial overlap entirely and even the off-diagonal peaks of $\rho^{(2)}$ become isolated. We term this behavior the formation of an off-diagonal correlation hole. For crystallized bosons, the spread of the anti-diagonal of the two-body density, $\rho^{(2)}(x,-x)$, is diverging as $g_d$ is increasing [compare Fig.~\ref{Fig2}(b) for $g_d=10$ to Fig.~\ref{Fig2}(b) for $g_d=30$]. 

We assert that {\it {the correlation hole along the diagonal and the off-diagonal}} and {\it {the unbounded spreading}} are the unique signatures of the two-body density of a crystalline state of dipolar bosons.

\subsection{Sorting Crystallization from Fermionization}\label{sec:CXF}

We now discuss how to distinguish fermionized from crystallized many-body states. One clear distinction is given by the spread of the one- and two-body densities: for bosons with contact interactions it is bounded, while for bosons with dipole-dipole interactions it diverges as a function of the interaction strength. 
We assert that, {\bf 1)} the bounded spreading of the density for contact interactions is a consequence of the bounded energy as the interaction strength tends to infinity. 
Similarly, we assert, {\bf 2)} that the unbounded spreading of the density for dipole-dipole interactions is a consequence of the unbounded energy as the interaction strength $g_d$ tends to infinity.
To validate the assertions {\bf 1)} \& {\bf 2)}, we quantify the spreading of the density as a function of interactions and plot the position of its outermost peak as a function of the interaction strength in Fig.~\ref{Fig3}(a) and compare it to the energy in Fig.~\ref{Fig3}(b), for $N=4$.

From fitting the energy in Fig.~\ref{Fig3}(b) we can infer that the energy as a function of contact interaction strength approaches the fermionization limit exponentially, a power law does not fit as accurately the data. For very large interactions, in the limit of $\lambda^{-1}\rightarrow 0$, our results are in agreement with the analysis in Ref.~\cite{Zinner:14}, see Appendix~\ref{diffN}. For dipolar interactions, the growth of the energy as a function of 
the interaction strength is fitting well to a power law.

Indeed, the comparison of Fig.~\ref{Fig3}(a) and Fig.~\ref{Fig3}(b) corroborates our assertions {\bf 1)} \& {\bf 2)}, and holds for different number of particles. 

We thus conclude that the crystalline phase can be distinguished from the Tonks gas by virtue of the behavior of its density profile as a function of the strength of the interparticle interactions: The width of the density distribution \textit{converges} for an increasing strength of contact interactions, but it \textit{continuously spreads} for an increasing strength of long-range interactions [compare Fig.~\ref{Fig1}(b) with Fig.~\ref{Fig1}(d) as well as Fig.~\ref{Fig3}(a) with Fig.~\ref{Fig3}(b)]. In Appendix~\ref{diffN}, we demonstrate that the exponent of the power law of the spreading of the density as a function of the strength of the interaction is independent of the particle number $N$.

In the case of long-ranged interactions, the unbounded spreading of densities as a function of increasing interaction strength and the formation of well-isolated peaks are in sharp contrast to the bounded spreading of densities and the non-isolated peaks in the case of contact interactions in the TG regime [cf. Fig.~\ref{Fig1} and Fig.~\ref{Fig3}(a)--(b)].

We now turn to analyze the eigenvalues of the reduced one-body density matrix, the so-called natural occupations~\cite{Penrose:56}, as a function of the interaction strength between the particles [Fig.~\ref{Fig3}(c)--(d)]. As expected~\cite{Zollner:06,Chatterjee:17,Chatterjee:12}, when the value of the interaction strength increases, the occupation of the first natural orbital decreases while the other orbitals start to be occupied.
For contact interactions, mostly one natural occupation, $n_1$, dominates, while the other occupations $n_k,k>1$ remain comparatively small even for large values of $\lambda$: depletion emerges as the fermionized state is reached [Fig.~\ref{Fig3}(c)], cf. also Ref.~\cite{Zollner:06}.
For long-range interactions, however, all occupations $\rho_k^{(NO)}$ for $k\leq N$ contribute on an equal footing for large values of $g_d$. This full-blown $N$-fold fragmentation emerges as the crystal state is reached [Fig.~\ref{Fig3}(d)], see also Ref.~\cite{Chatterjee:17}. In the crystal state the bosons behave similar to distinguishable particles \cite{Kocik:18}, and the particle statistics does not influence the physical observables considered in Fig.~\ref{Fig3}(b)(d): the energy and the natural occupations for bosons and fermions converge to the same values. Thus, the finding of Ref. \cite{Kocik:18} for two particles may be extended to larger number of particles.      

The emergence of complete fragmentation is a consequence of long-ranged interactions and in sharp contrast to the emergent depletion in the case of contact interactions. 
 
 \section{Conclusions}\label{sec:conclusions}
 
In this paper we highlight the key characteristics of the many-body wavefunction that reveal the difference between the fermionized bosons with contact interactions and crystallized bosons with dipolar 
interactions. 

In the case of fermionization, the one-(two-)body density shows a modulation with a number of maxima corresponding to the number of particles. The maxima are confined but not completely separated. The incomplete separation is a consequence of the representability of momentum distribution of fermionized bosons using a basis set: infinitely many basis states are necessary to accurately resolve the cusp -- a fact that is reflected by the depletion of the state which we quantified by the eigenvalues of the reduced one-body density matrix. We found that the peaks in the density as well as the energy as a function of the interaction strength approach the fermionization limit exponentially.

In the case of crystallization, the one-(two-)body density shows well-separated peaks whose distances diverge as a function of the interaction strength as a power law. This completed separation is the consequence of the formation of a Mott-insulator-alike many-body state where the ``lattice potential'' is replaced by the long-ranged interparticle interactions and the ``lattice constant'' is dictated by the strength of the interparticle interactions. 

We close by stating that all the signatures that distinguish crystalline bosons from fermionized bosons can be measured experimentally using single-shot absorption imaging~\cite{Sakmann:16,Javanainen:96,Castin:97,Dziarmaga:03,Dagnino:09}. From experimental absorption images, the one-body and two-body density are available as averages of many single-shot images. Thus, a direct verification of our results for the spread of the one-body and two-body density can be performed. 
Furthermore, Refs.~\cite{Chatterjee:17,Lode:17} suggest that the natural occupations can be inferred from the integrated variance of single-shot images, at least at zero temperature. It is, of course, an open question how thermal fluctuations affect the variance in absorption images and up to which temperature it is still possible to determine the fragmentation of the system.

\begin{acknowledgments}
S. Bera acknowledges DST (Govt. of India) for the financial support through INSPIRE fellowship [2015/IF150245] and Rhombik Roy for some useful discussions. B. Chakrabarti acknowledges FAPESP (grant No. 2016/19622-0). AG and MCT acknowledge FAPESP and AG thanks CNPq for financial support. B. Chatterjee acknowledges financial support from the Department of Science and Technology, Government of India under the DST Inspire Faculty fellowship. AUJL and CL acknowledge financial support by the Austrian Science Foundation (FWF) under grant Nos. P 32033 and M 2653, respectively, and the Wiener Wissenschafts- und TechnologieFonds (WWTF) project No MA16-066 . 
\vspace*{2cm}
\end{acknowledgments}

\section*{Contributions}

S.~B. conducted the numerical and theoretical investigations, S.~B., C.~L. and A.~U.~J.~L. wrote the manuscript, A.~U.~J.~L. and C.~L. performed complementary calculations, S.~B., B.~Chak., A.~G., M.~C.~T., M.~L.~L., B.~Chat., C.~L., and A.~U.~J.~L. conceived the idea for the project and supported the writeup.

\section*{Competing Interests}
The authors declare no competing interests.

\onecolumngrid

\begin{figure}[htbp]
\includegraphics[height=\textwidth,angle=-90]{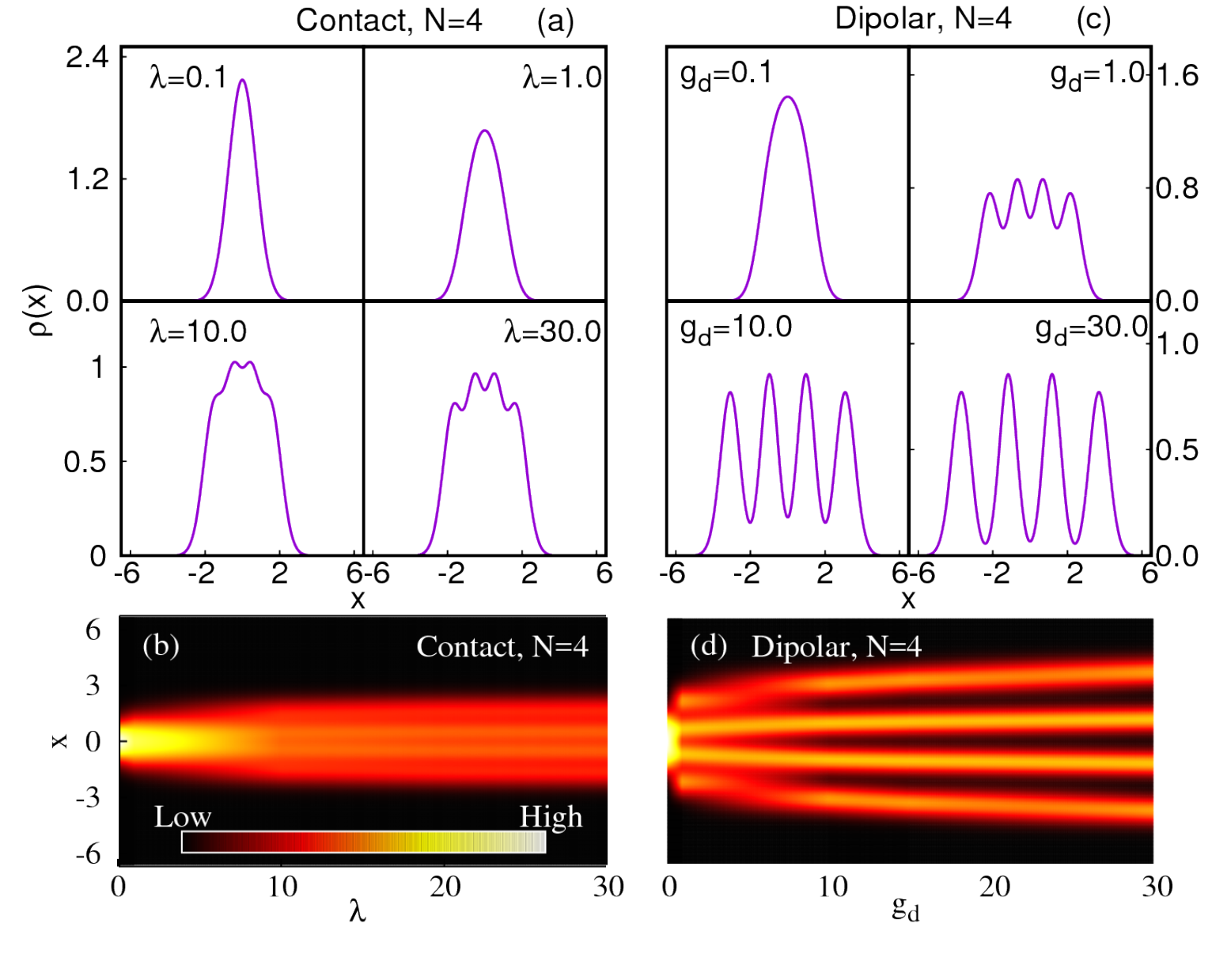}
\caption{One-body density of four bosons as a function of contact [(a),(b)] and dipolar [(c),(d)] interparticle interaction strength. \textit{For contact interactions}, the density becomes flatter and broader as the repulsion increases [panel (a) and (b) for $\lambda\leq1$]. For even larger interaction strengths [panel (a) and (b) for $\lambda\gtrsim10$], four distinct but not isolated peaks appear and the density gradually converges to the density of four non-interacting fermions as $\lambda\rightarrow\infty$. Due to this convergence, the spread of the density seizes to increase [panel (d)].
\textit{For dipolar interactions}, the one-body density is clustered at the center of the trap for small interactions [panels (c),(d) for $g_d\lesssim1$]. As $g_d$ increases, the density develops a fourfold splitting [panel (c) and (d) for $g_d\gtrsim1$]. As a function of increasing interaction strength, the spread of the density continues to increase [panel (d)] and the fourfold spatial splitting intensifies to form four completely isolated peaks in the density for sufficiently strong dipolar interactions: crystallization emerges [panels (c),(d) for $g_d\gtrsim10$]. All quantities shown are dimensionless.}
\label{Fig1}
\end{figure}

\clearpage
\twocolumngrid

\begin{figure}
	\includegraphics[height=0.5\textwidth,angle=-90]{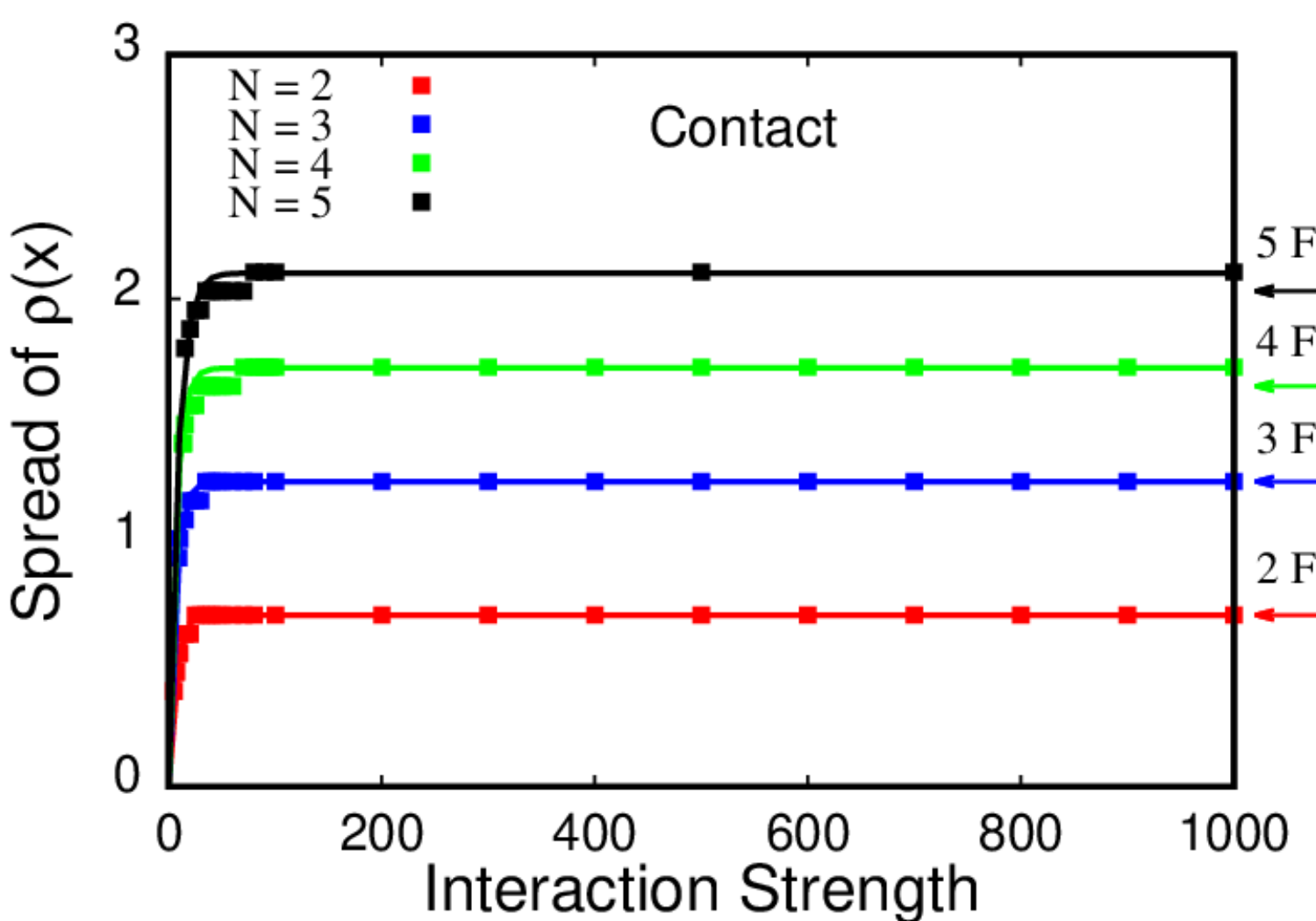}
	\caption{Spread of the density $\rho(x)$ as a function of the interaction strength for $N=2,3,4,5$ (bottom to top curve, respectively) bosons  with contact interparticle interactions. The spread of the density, according to the fitted curves (solid lines) converges exponentially as $A_N (\exp{-\lambda/B_N}-1)$ to the fermionization limit as $\lambda \rightarrow \infty$ which is shown by the arrows labeled ``2F'', ``3F'', ``4F'', ``5F'' on the right hand side of the plot. The fit parameters for $N=2,3,4,5$ are, respectively, $(A_2=-0.701491,B_2=6.45191),(A_3=-1.25018,B_3=6.50518),(A_4=-1.71554,B_4=6.8185),(A_5=-2.10423,B_5=8.63662)$. All quantities shown are dimensionless.}
	\label{Fig5}
\end{figure}

\clearpage
\onecolumngrid

\begin{figure}
\includegraphics[width=\textwidth]{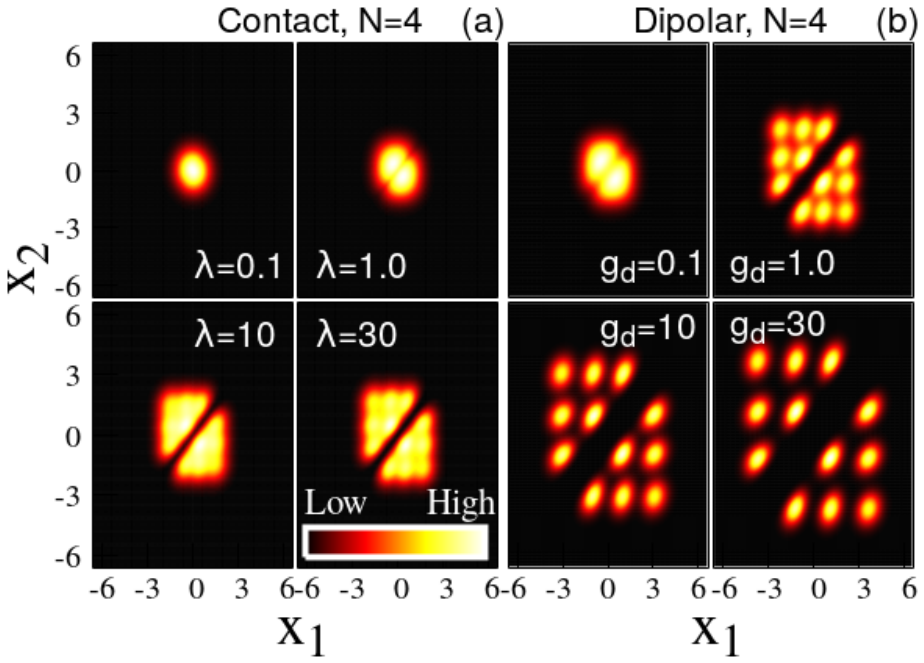}
\caption{Two-body density of four bosons as a function of contact (a) and dipolar [(b)] interparticle interaction strength.
\textit{For contact interactions}, the atoms are clustered at the center ($x_1=x_2=0$) for small interaction strengths, [panel (a) for $\lambda=0.1$]. As $\lambda$ increases, the two-body density starts to spread due to the repulsion between the bosons [panel (a) for $\lambda=1$]. For stronger interaction strengths, $\lambda=10$ and $\lambda=30$ in (a), the diagonal, $\rho^{(2)}(x,x)$, is practically $0$: the bosons completely avoid to be at the same position and a ``correlation hole'' develops. \textit{For dipolar interactions}, the atoms cluster at the center ($x_1=x_2=0$) for small interaction strengths, see panel (b) for $g_d=0.1$. As $g_d$ increases, the diagonal part, $\rho^{(2)}(x,x)$ starts to be depleted because the long-range interactions start to dominate the physics [panel (b) for $\gtrsim1$]. At stronger interaction strengths, the diagonal correlation hole spreads, i.e., the area in the vicinity of $x_1\approx x_2$ for which $\rho^{(2)}(x_1,x_2)\approx 0$ holds is enlarged as a function of $g_d$ [compare panel (b) for $g_d=1.0,10,$ and $30$]. 
In contrast to contact interactions, even the off-diagonal ($x_1\neq x_2$) of $\rho^{(2)}(x_1,x_2)$ forms a complete correlation hole, compare panel (a) for $\lambda=30$ and panel (b) for $g_d=30$. All quantities shown are dimensionless.}
\label{Fig2}
\end{figure}

\clearpage
\twocolumngrid

\begin{figure}
 \includegraphics[height=0.5\textwidth,angle=-90]{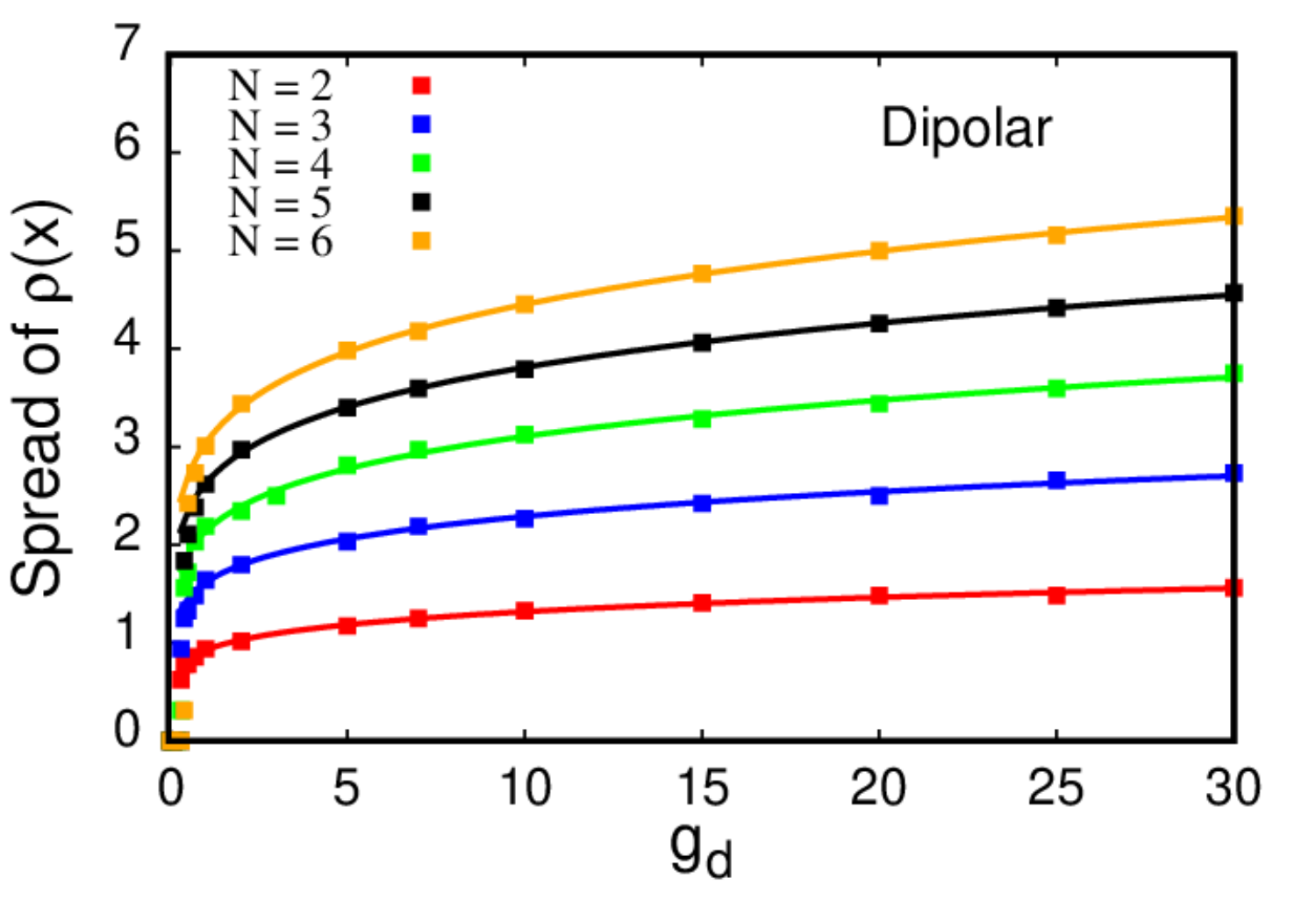}
\caption{Spread of the density for $N=2,3,4,5,6$ bosons (bottom to top curve, respectively) with dipole-dipole interactions as a function of the interaction strength $g_d$. The spread of the density, according to the fitted curves (solid lines) diverges as a power law, $C_N x^{D_N}$, in the limit of large interactions $g_d \rightarrow \infty$. The fit parameters for $N=2,3,4,5,6$ are, respectively, $(C_2= 0.926851, D_2=0.152459),(C_3=1.61556,D_3=0.151243),(C_4=2.13826, D_4=0.162034),(C_5=2.62615,D_5=0.161553),(C_6=3.03802,D_6=0.165883)$. Importantly, the power of the divergence of the spread, $D_N$, seems to be independent of the number of particles $N$. All quantities shown are dimensionless.}
\label{Fig8}
\end{figure}

\begin{figure}
\begin{center}
 \includegraphics[height=.45\textwidth,angle=-90]{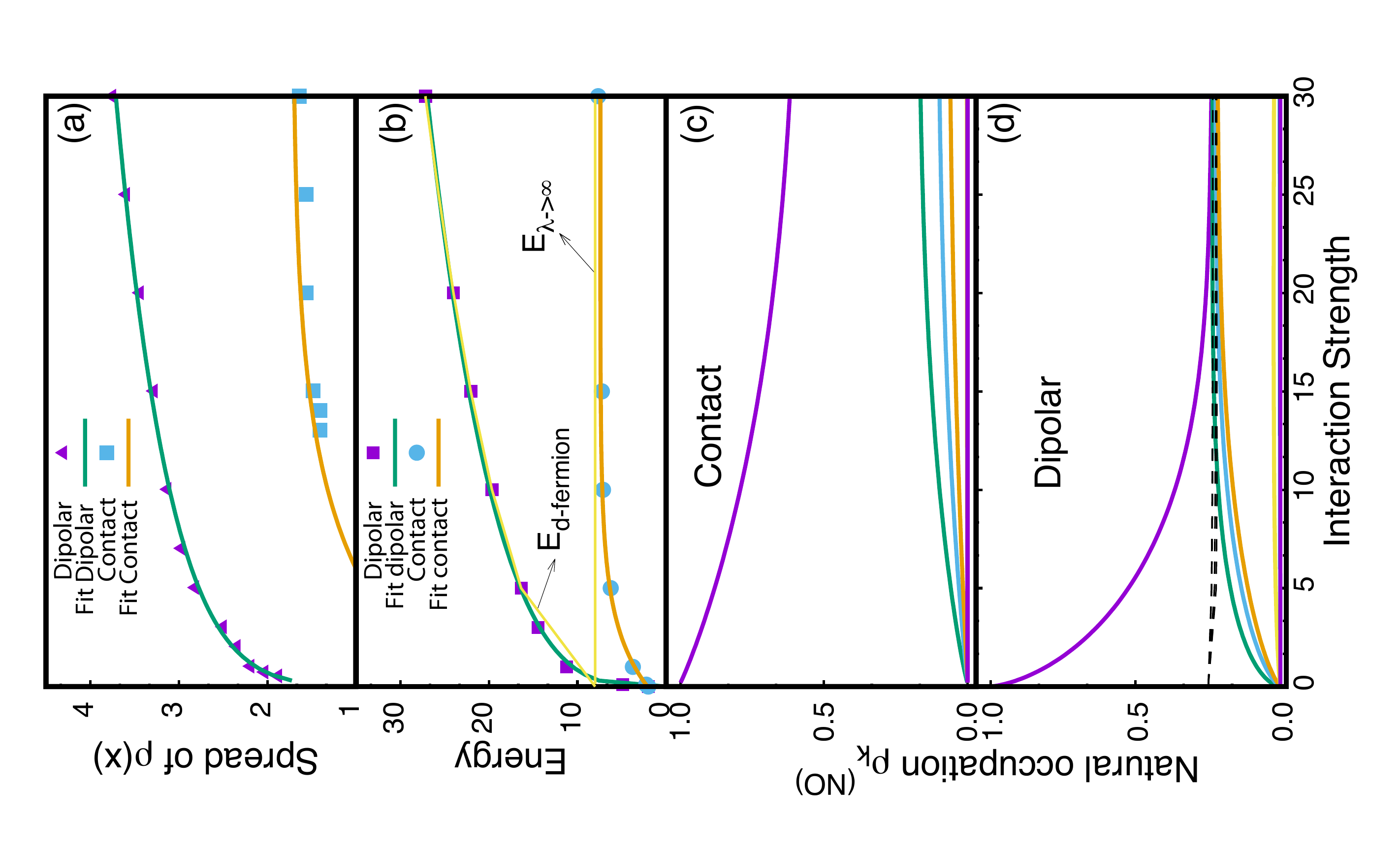}
\end{center}

\caption{Tracing fermionization and crystallization in the spread of the density (a), the energy (b), and the natural occupations (c),(d) as a function of the interaction strength. (a) The spread of the density is quantified by the position of the outermost peak in the density $\rho(x)$. The spread is bounded for contact interactions and unbounded for dipolar interactions. The fits shown suggest that the spread of the density $\rho(x)$ for dipolar interactions diverges with a power law,  $2.138 g_d^{0.162}$, and for contact interactions it converges as $-1.71554 (\exp{(-\lambda/6.8185)}-1)$ to the fermionization limit (fit obtained with more points than actually shown, see Appendix~\ref{diffN}).
(b) The energy as a function of interaction strength is bounded for contact interactions and unbounded for dipolar interactions. The fits suggest that the energy diverges with a power law $10.51 g_d^{0.277}$ for dipolar interactions and converges to the fermionization limit exponentially $-5.84\exp{(-\lambda/6.023)}+8.133$ for contact interactions.
The thin yellow lines indicate the energy of non-interaction fermions $E_{\lambda\rightarrow\infty}$ and the energy for dipolar fermions $E_{d-fermion}$. (c),(d) the eigenvalues of the reduced density matrix, i.e., the natural occupations $\rho_i^{(NO)}$ exhibit depletion for contact interactions (many small $\rho_i^{(NO)}$ with $i>1$ emerge) and full-blown $N$-fold fragmentation for dipolar interactions (all $\rho_i^{(NO)}$ with $i\leq N$ contribute equally), the black dashed lines show the four most populated natural orbitals for dipolar fermions. The $\rho_i^{(NO)}$ are ordered in decreasing order starting from $i=1$. All quantities shown are dimensionless.}
\label{Fig3}
\end{figure}

\clearpage

\appendix
\section{Comparison of MCTDHB and exact diagonalization}\label{MCTDHBvsED}
\begin{figure}
	\includegraphics[height=0.5\textwidth,angle=-90]{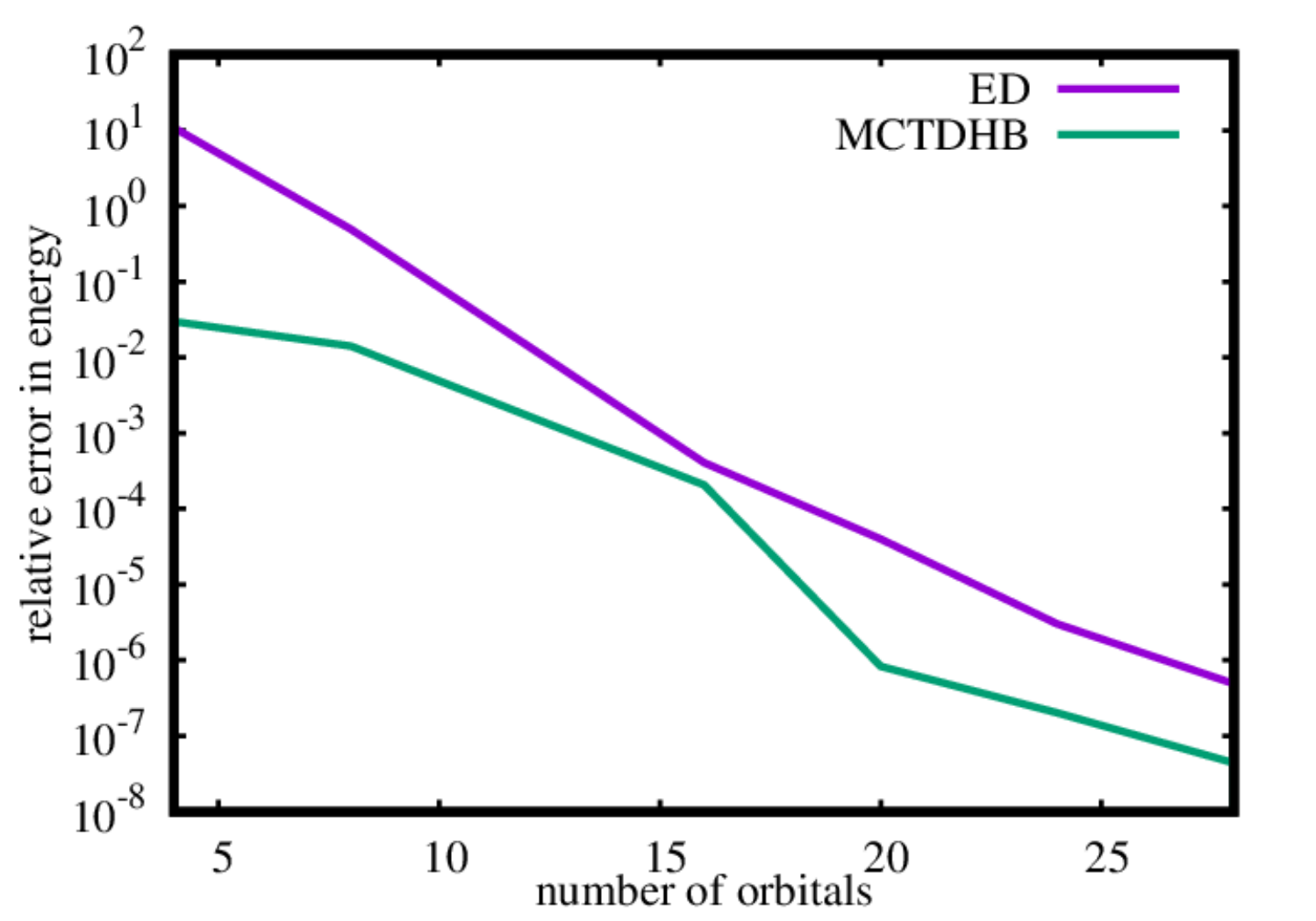}
	\caption{Comparison of MCTDHB and ED for $N=4$ bosons with dipole-dipole interaction strength $g_d=30$. The plot shows the relative error in energy  with respect to an MCTDHB computation with $M=32$ orbitals as a function of the number of orbitals for the ED and MCTDHB approaches. Due to the variationally optimized basis in MCTDHB computations it features a much smaller error for any number of orbitals. All quantities shown are dimensionless.}
	\label{Fig4}
\end{figure}

In this Appendix, we demonstrate that MCTDHB yields solutions to the Schrödinger equation at a larger accuracy as compared to the exact diagonalization approach (ED). As is conventional, we use the eigenfunctions of the non-interacting system as the single-particle basis states for the ED. We solve the same system as shown in Fig.~\ref{Fig1}c) and d) for an interaction strength of $g_d=30$ and compare the energies obtained with MCTDHB and ED, see Fig.~\ref{Fig4}. Due to the variationally optimized single-particle basis in MCTDHB computations it features a much smaller error than the ED computations with an unoptimized single-particle basis for the same number of orbitals. 
This observation is in agreement with other works that benchmark the MCTDHB and the MCTDHF approaches against ED, see Ref.~\cite{Axel:HIM} and Ref.~\cite{MCTDHF}, respectively.

\section{Different particle numbers}\label{diffN}
In this Appendix, we corroborate our results in the main text by studying different particle numbers. 

\subsection{Contact interactions}
The results of the manuscript have been obtained with MCTDHB with $M=12,14,20,22$ orbitals for $N=2,3,4,5$ bosons, respectively, with a contact interaction strength up to $\lambda=1000$. In Fig.~\ref{Fig5} of the main text, our results are consistent with fits of an exponential function $A_N (\exp(-\lambda/B_N) -1)$, see caption of Fig.~\ref{Fig5} for the fitting parameters $A_N$ and $B_N$. Furthermore, we assess the convergence of the spread of the density as a function of the interaction strength to the spread of the density of the non-interacting fermionic system, see arrows labeled ``2F'',``3F'',``4F'', and ``5F'' in Fig.~\ref{Fig5}.

\begin{figure}
 \includegraphics[height=0.5\textwidth,angle=-90]{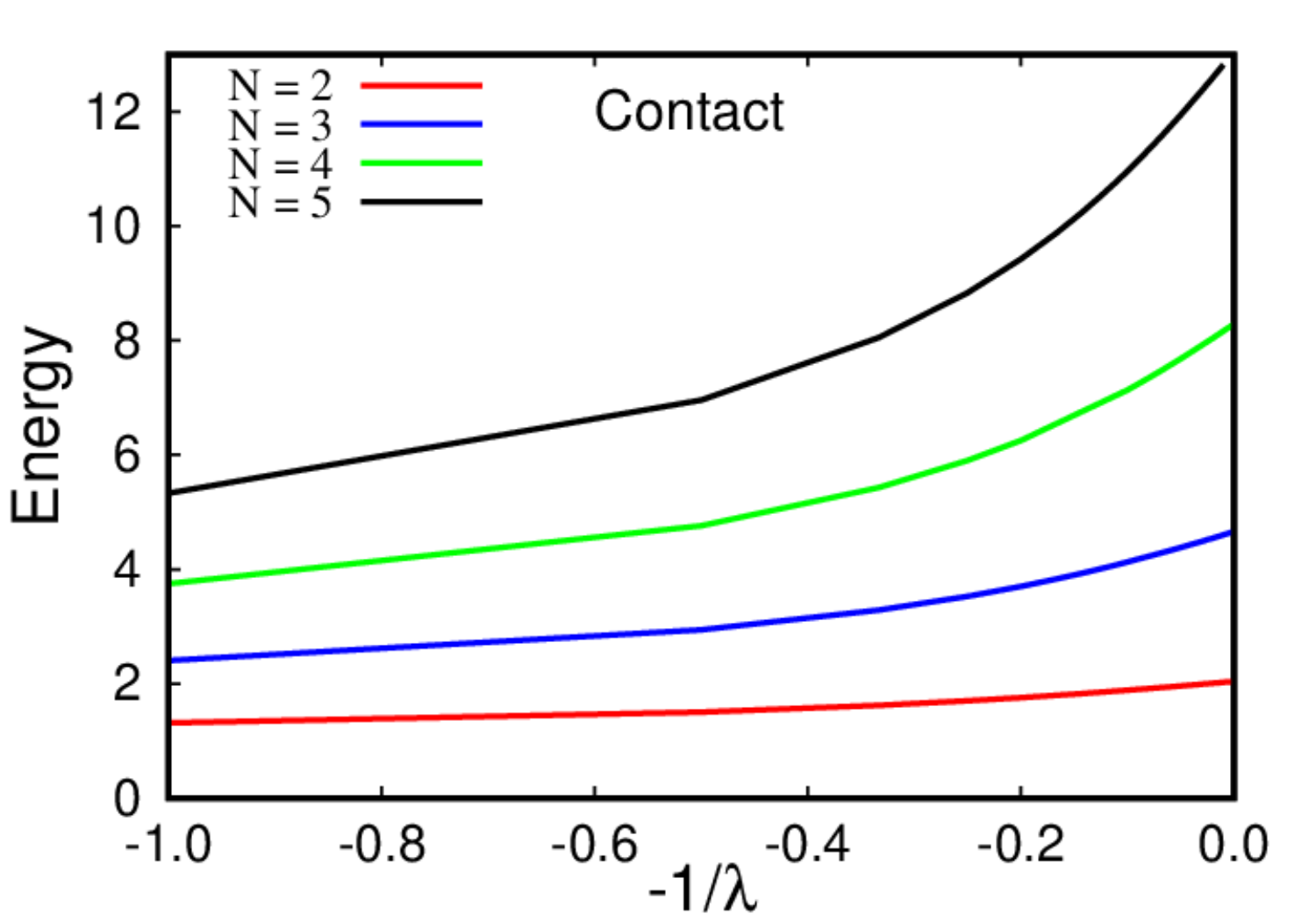}
\caption{Energy as a function of inverse interaction strength, $-1/\lambda$, for $N=2,3,4,5$ (bottom to top curve, respectively) bosons. Our results are consistent with the analysis in Ref.~\cite{Zinner:14}: the energy linearly converges to the fermionization limit, i.e., when $-\lambda^{-1} \rightarrow 0$. All quantities shown are dimensionless.}
\label{Fig6}
\end{figure}

To compare our results for the energy in the fermionization limit to analytical predictions for very large contact interaction strengths in Ref.~\cite{Zinner:14}, we plot the energies as a function of $-\lambda^{-1}$ in Fig.~\ref{Fig6}. We find that our results are consistent with the linear limit for the energy as a function of $-\lambda^{-1}$ of Ref.~\cite{Zinner:14}.

We now turn to the relative height of the innermost and outermost peak(s), 
\begin{equation}
 \Delta\rho(x) = \frac{\rho_{max}-\rho_{min}}{\rho_{max}+\rho_{min}}\label{RPH}
\end{equation}
in the density. Here, $\rho_{max}$ refers to the value of the density $\rho(x)$ at the peak position and $\rho_{min}$ refers to the value of the density $\rho(x)$ at the position of the minimum to the left to the considered peak. See Fig.~\ref{Fig7} for a plot of $\Delta\rho(x)$ for $N=2,3,4,5$ bosons. It is clearly seen that, for fixed $N$, the outermost peaks' relative height is much smaller than the relative height of the innermost peaks. 

\subsection{Dipolar interactions}

Here, we assess the validity of the power-law-like unbounded spreading of the density as a function of the strength of dipole-dipole interactions, that we have shown in Fig.~\ref{Fig8} of the main text for $N=4$ particles. In Fig.~\ref{Fig8} We plot the spread of the density for $N=2,3,4,5,6$ dipolar bosons obtained with MCTDHB with $M=16,16,22,28,26$ orbitals, respectively, and fit it with a power law $C_N g_d^{D_N}$. We find that the exponent in the power law is almost identical for all particle numbers studied here, i.e., $D_N\approx 0.16$ for $N=2,3,4,5,6$.

\begin{figure}
 \includegraphics[height=0.5\textwidth,angle=-90]{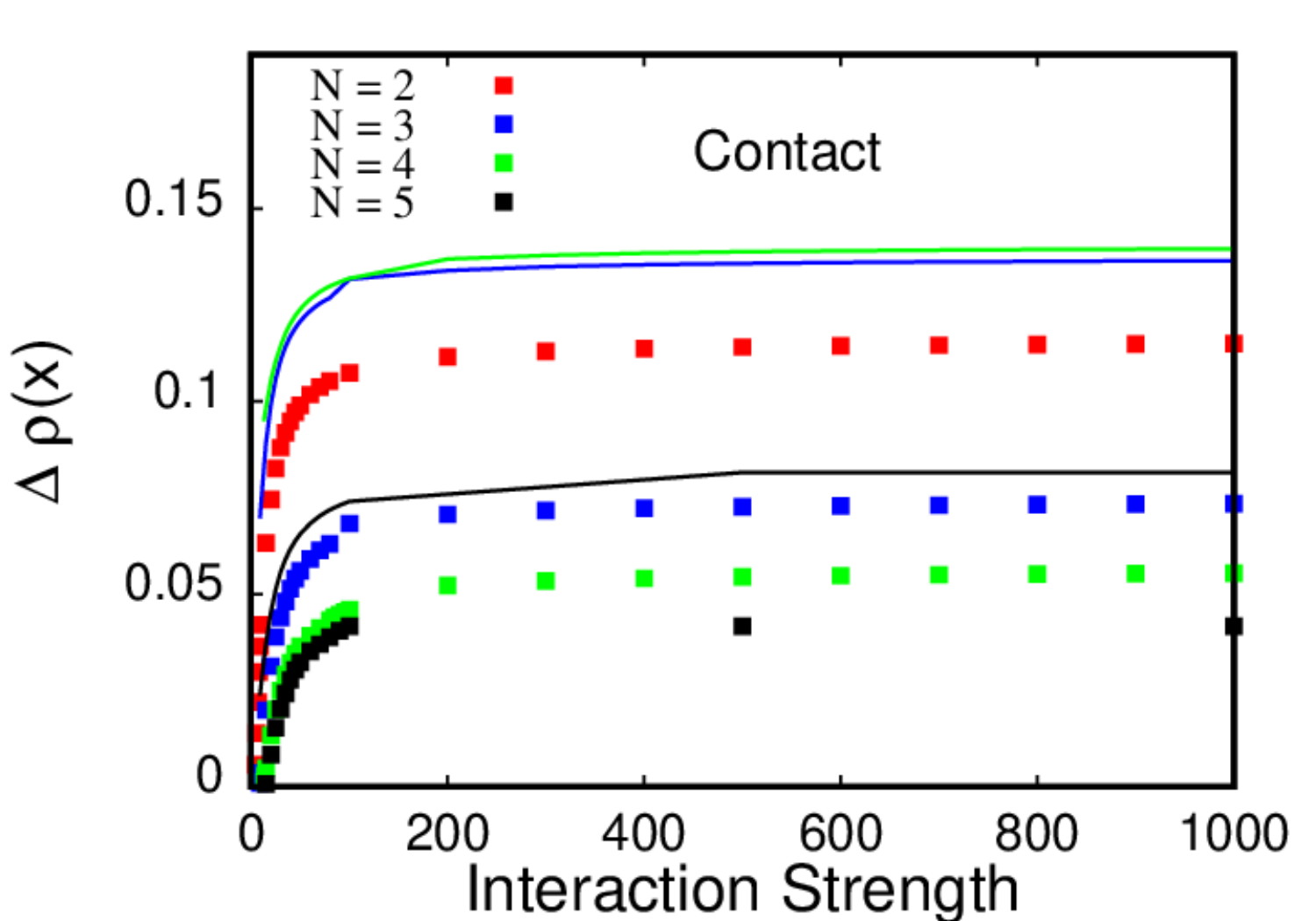}
\caption{Relative height of the outermost peaks in the density of $N=2,3,4,5$ bosons with contact interactions (points, top to bottom, respectively) and relative height of the innermost peaks in the density of $N=4,3,5$ bosons with contact interactions (lines, top, to bottom, respectively). The relative peak height is consistently smaller for the outermost peak as compared to the innermost peak in the density for all interaction strengths depicted. All quantities shown are dimensionless.}
\label{Fig7}
\end{figure}

\begin{figure}
 \includegraphics[height=0.5\textwidth,angle=-90]{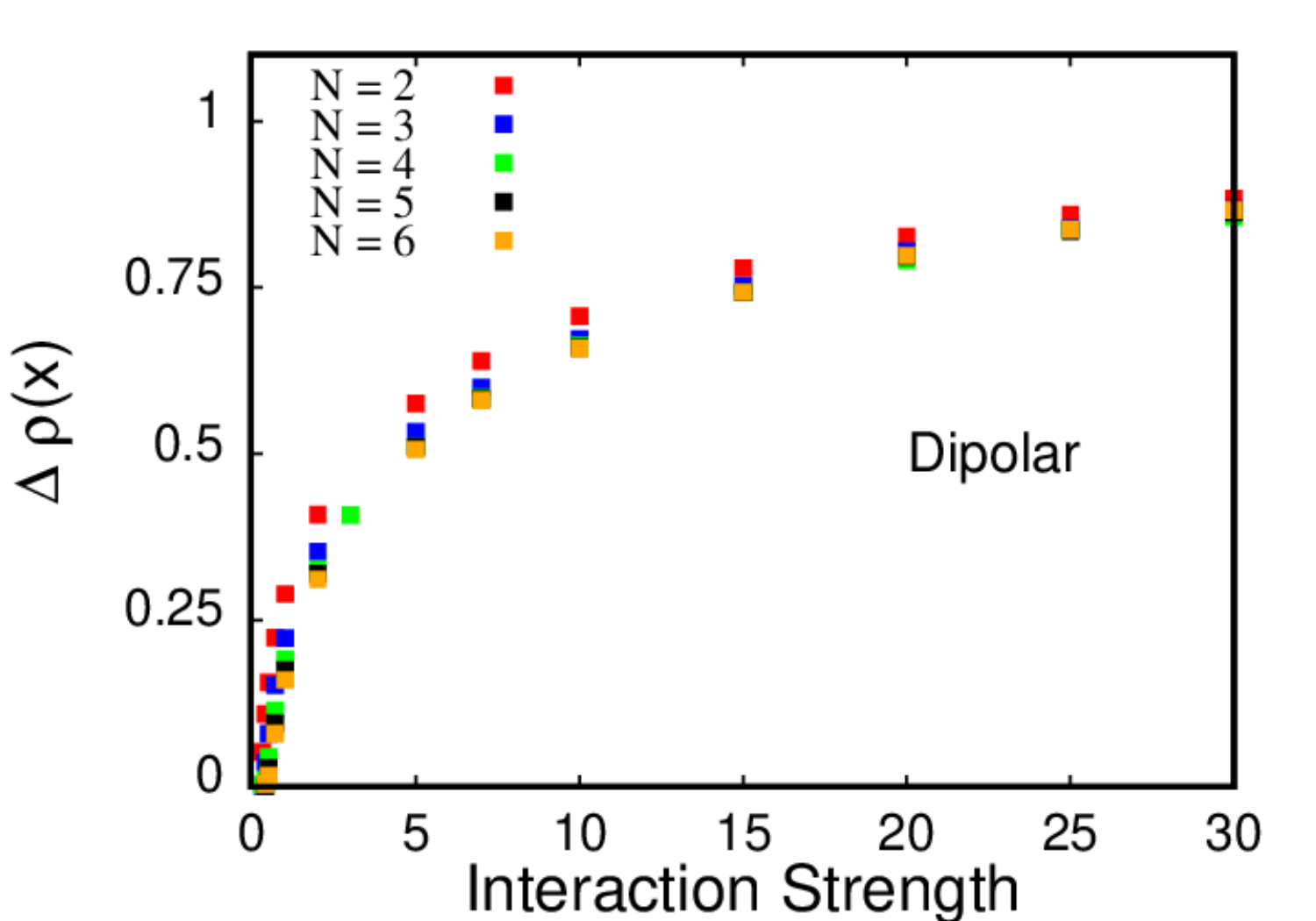}
\caption{Relative peak height for $N=2,3,4,5,6$ bosons with dipolar interactions as a function of interaction strength. The relative peak height converges towards unity similarly for all particle numbers investigated here as the interaction strength increases. All quantities shown are dimensionless.}
\label{Fig9}
\end{figure}

We now discuss the relative peak height $\Delta\rho(x)$, see Eq.~\eqref{RPH}, of the outermost peak as a function of the dipolar interaction strength, see Fig.~\ref{Fig9} for a plot for $N=2,3,4,5,6$. As hinted by Fig.~\ref{Fig1}c) in the main text, the relative peak height for the case of the dipole-dipole interactions converges towards unity as the strength of interactions $g_d$ increases, because the values of the minimum, $\rho_{min}$ in Eq.~\eqref{RPH}, tends to zero: the peaks in the crystal state are {\it well-isolated} in comparison to the peaks in the fermionization limit for bosons with contact interactions (compare magnitude of relative peak heights in Figs.~\ref{Fig7} and \ref{Fig9}).

\clearpage


\begin{thebibliography}{999}
\bibitem{Petrov:00} D. S. Petrov, G. V. Shlyapnikov, and J. T. M. Walraven, Phys. Rev. Lett. {\bf 85}, 3745 (2000).
\bibitem{Dunjko:01} V. Dunjko, V. Lorent, and M. Olshanii, Phys. Rev. Lett. {\bf 86}, 5413 (2001).
\bibitem{Gorlitz:01} A. G\"orlitz {\it et al.}, Phys. Rev. Lett. {\bf 87}, 130402 (2001).
\bibitem{Greiner:01} M. Greiner {\it et al.}, Phys. Rev. Lett. {\bf 87}, 160405 (2001).

%%% Fermionization EXP
\bibitem{Zurn:12} G. Zürn, F. Serwane, T. Lompe, A. N. Wenz, M. G. Ries, J. E. Bohn, and S. Jochim
Phys. Rev. Lett. {\bf 108}, 075303 (2012).
\bibitem{Jacqmin:11} T. Jacqmin, J. Armijo, T. Berrada, K.~V. Kheruntsyan, and I. Bouchoule
Phys. Rev. Lett. {\bf 106}, 230405 (2011).

%%% Fermionization Theory
\bibitem{Girardeau:60} M. Girardeau, J. Math Phys. {\bf{1}}, 516 (1960).
\bibitem{Zollner:06} S. Zöllner, H.-D. Meyer, and P. Schmelcher, Phys. Rev. A {\bf 74}, 063611 (2006). 
\bibitem{Alon:05} O.~E. Alon and L.~S. Cederbaum, Phys. Rev. Lett. {\bf 95}, 140402 (2005).

%%% Fermionization beyond single-component systems
\bibitem{Zollner:08} S. Zöllner, H.-D. Meyer, and P. Schmelcher, Phys. Rev. A {\bf 78}, 013629 (2008).
\bibitem{Roy:18} R. Roy, A. Gammal, M. C. Tsatsos, B. Chatterjee, B. Chakrabarti, and A. U. J. Lode, Phys. Rev. A {\bf 97}, 043625 (2018).

 \bibitem{Kocik:17b}  P. Ko\'scik, Few-Body Syst. {\bf 58}, 59 (2017)

% crystallization
 \bibitem{arkhipov05}
 A.~S. {Arkhipov}, G.~E. {Astrakharchik}, A.~V. {Belikov}, and Y.~E. {Lozovik}, JETP Lett. {\bf 82},  39  (2005).
  \bibitem{Zollner:11pra}
S. Z\"ollner, G.~M. Bruun, C.~J. Pethick, and S.~M. Reimann, Phys. Rev. Lett. \textbf{107}, 035301 (2011).
\bibitem{Zollner:11} S. Z\"ollner, Phys. Rev. A {\bf 84}, 063619 (2011)
\bibitem{astrakharchik:08a} G.~E. Astrakharchik, G.~E. Morigi, G. De~Chiara, and J. Boronat, Phys. Rev. A {\bf78},  063622  (2008).
\bibitem{astrakharchik:08b} G.~E. Astrakharchik and Yu.~E. Lozovik, Phys. Rev. A {\bf77}, 013404 (2008).
\bibitem{Deuretzbacher:10} F. Deuretzbacher, J. C. Cremon, and S. M. Reimann, Phys. Rev. A {\bf 81}, 063616 (2010).

\bibitem{Sowinski:19} T. Sowi\'nski and M. \'A. Garc\'ia-March,  arXiv:1903.12189 [cond-mat.quant-gas]



\bibitem{Paredes:04} B. Paredes {\it et al.}, Nature {\bf{429}}, 277 (2004).
\bibitem{Deuretzbacher:07} F. Deuretzbacher {\it et al.}, Phys. Rev. A {\bf 75}, 013614 (2007).
\bibitem{Kinoshita:04} T. Kinoshita, T.Wenger, and D. S. Weiss, Science {\bf{ 305}}, 1125 (2004).
\bibitem{Santos:03} L. Santos, G. V. Shlyapnikov, and M. Lewenstein, Phys. Rev. Lett. {\bf 90}, 250403 (2003).
\bibitem{Andre:06} A. Andre {\it et al.}, Nature Phys. {\bf 2}, 636 (2006).
\bibitem{Buchler:07} H. P. Buchler {\it et al.}, Phys. Rev. Lett. {\bf 98}, 060404 (2007).

% experiments with DD

% The physics of dipolar bosonic quantum gases
\bibitem{baranov08} M.~A. {Baranov}, Phys. Rep. {\bf 464},  71  (2008).
\bibitem{griesmaier05} A. Griesmaier, J. Werner, S. Hensler, J. Stuhler, T. Pfau, Phys. Rev. Lett. \textbf{94}, 160401 (2005).
\bibitem{beaufils08} Q. Beaufils \emph{et al.}, Phys. Rev. A \textbf{77}, 061601 (2008).

\bibitem{Lahaye:09} T. Lahaye, C. Menotti, L. Santos, M. Lewenstein, and T. Pfau,
Rep. Prog. Phys. {\bf 72}, 126401 (2009).

\bibitem{Romanovsky:04} I. Romanovsky, C. Yannouleas, and U. Landman, Phys. Rev. Lett. {\bf 23}, 230405 (2004)
 \bibitem{Kocik:17}  P. Ko\'scik, Eur. Phys. J. D {\bf 71}, 286 (2017).
 \bibitem{Kocik:15}  P. Ko\'scik, Phys. Lett. A {\bf 379}, 293 (2015).

\bibitem{Chatterjee:12} B. Chatterjee, I. Brouzos, L. Cao, and P. Schmelcher, J. Phys. B: At. Mol. Opt. Phys. {\bf 46}, 085304 (2013).
\bibitem{Chatterjee:17} B. Chatterjee and A.~U.~J. Lode, Phys. Rev. A {\bf 98}, 053624 (2018).

\bibitem{Wigner} E. Wigner, Phys. Rev. {\bf 46}, 1002 (1934).

\bibitem{Ferlaino:12} K. Aikawa, A. Frisch, M. Mark, S. Baier, A. Rietzler, R. Grimm, and F. Ferlaino, Phys. Rev. Lett. {\bf 108}, 210401 (2012).
\bibitem{Ferlaino:15} A. Frisch, M. Mark, K. Aikawa, S. Baier, R. Grimm, A. Petrov, S. Kotochigova, G. Qu\'em\'ener, M. Lepers, O. Dulieu, and F. Ferlaino, Phys. Rev. Lett. {\bf 115}, 203201 (2015).

\bibitem{Zwierlein:15} Jee Woo Park, Sebastian A. Will, and Martin W. Zwierlein, Phys. Rev. Lett. {\bf 114}, 205302 (2015).


\bibitem{CONDMAT} Z. Xu,  L. Li, G. Xianlong, and S. Chen, J. Phys.: Condens. Matter {\bf 25} 055601 (2013).

%%% GP 
\bibitem{Pethick:02} C.\,J. Pethick and H. Smith, {\em Bose-Einstein Condensation in Dilute Gases} (Cambridge University Press, Cambridge UK, 2002).

\bibitem{Bogoliubov:91} N.\,N. Bogoliubov, {\em Selected Works %, Part 
II: Quantum and Statistical Mechanics} (Gordon and Breach, New York, 1991). 

\bibitem{Pitaevskii:03} L.\,P. Pitaevskii and S. Stringari,   {\em Bose-Einstein Condensation} (Clarendon Press, Oxford, 2003).
  
 % CI like calculation
 \bibitem{MCTDHB1}  Ofir E. Alon, Alexej I. Streltsov, and Lorenz S. Cederbaum, 
Phys. Rev. A {\bf 77}, 033613 (2008).
\bibitem{MCTDHB2} A. I. Streltsov, O. E. Alon, and L. S. Cederbaum, Phys. Rev. Lett. {\bf 99}, 030402 (2007).
\bibitem{Cao:13} L. Cao, S. Kr\"onke, O. Vendrell, P. Schmelcher, J. Chem. Phys.  {\bf 139}, 134103 (2013).
\bibitem{MCTDHF} E. Fasshauer and A.~U.~J. Lode, Phys. Rev. A {\bf 93}, 033635 (2016).
\bibitem{Leveque:17} C. L\'ev\^eque, L. B. Madsen, New Journal of Physics {\bf 19}, 043007 (2017).
\bibitem{Leveque:18} C. L\'ev\^eque, L. B. Madsen, Journal of Physics B {\bf 51}, 155302 (2018).
\bibitem{Lode:19} A. U. J. Lode, C. L\'ev\^eque, L. B. Madsen, A. I. Streltsov and O. E. Alon, arXiv:1908.03578

\bibitem{Mese:11} A. I. Mese, P. Capuzzi, S. Aktas, Z. Akdeniz and S. E. Okan, Phys. Rev. A {\bf 84}, 043604 (2011).
\bibitem{Yannouleas:07} C. Yannouleas and U. Landman, Reports on Progress in Physics  {\bf 70}, 2067 (2007).


%%% MCTDHB


\bibitem{ultracold} A.\,U.\,J. Lode, M.\,C. Tsatsos, E. Fasshauer, R. Lin, L. Papariello, P. Molignini, and C. L\'ev\^eque, \textsc{MCTDH-X}:{\em The time-dependent multiconfigurational Hartree for indistinguishable particles software}, \href{http://ultracold.org}{http://ultracold.org} (2018).
\bibitem{Axel:Thesis} A.~U.~J. Lode, {\it Tunneling Dynamics in Open Ultracold Bosonic Systems}, Springer Theses, (Springer, Heidelberg, 2014).
\bibitem{Axel:Spin} A. U. J. Lode, Phys. Rev. A {\bf 93}, 063601 (2016).


\bibitem{Heimsoth:10} M. Heimsoth and M. Bonitz, Physica E {\bf 42}, 420 (2010).
\bibitem{units} We divide the dimensional Hamiltonian by $\hbar^2/(mL^2)$, where $m$ is the mass of the considered bosons and $L$ a conveniently chosen length scale.



\bibitem{Olshanii:98} M. Olshanii, Phys. Rev. Lett. {\bf 81}, 938 (1998).
\bibitem{Sinha:07} S. Sinha and L. Santos, Phys. Rev. Lett. {\bf 99}, 140406 (2007).
\bibitem{cai10} Y. Cai, M. Rosenkranz, Z. Lei and W. Bao, Phys. Rev. A {\bf 82}, 043623 (2010).

\bibitem{Penrose:56}O. Penrose and L. Onsager, 
%Bose-Einstein Condensation and Liquid Helium, 
Phys. Rev. {\bf 104}, 576 (1956).

\bibitem{Nozieres:82} P. Nozi\`{e}res, D. Saint James, J. Phys. (France) {\bf 43}, 1133 (1982).
\bibitem{Spekkens:99} R.\,W. Spekkens and J.\,E. Sipe, Phys. Rev. A {\bf 59}, 3868 (1999).

\bibitem{Lode:17} A. U. J. Lode and C. Bruder, Phys. Rev. Lett. {\bf 118}, 013603 (2017).

\bibitem{TDVP} {\it{Geometry of the time-dependent variational principle}}, P. Kramer and M. Saracen, (Springer, Berlin, 1981).

\bibitem{Axel:HIM} A. U. J. Lode, K. Sakmann, O. E. Alon, L. S. Cederbaum, and A. I. Streltsov, Phys. Rev. A {\bf 86}, 063606 (2012).
\bibitem{Gwak:18} Y. Gwak, O. V. Marchukov, U. R. Fischer,  arXiv:1811.04705 [cond-mat.quant-gas]

\bibitem{Zinner:14} N. T. Zinner, A. G. Volosniev, D. V. Fedorov, A. S. Jensen, and M. Valiente, Eur. Phys. Lett. {\bf 107}, 60003 (2014).

\bibitem{Kocik:18} P. Ko\'scik and T. Sowi\'nski, Scientific Reports {\bf 8}, 48 (2018). 
 

%Single shot
\bibitem{Sakmann:16} K. Sakmann and M. Kasevich, Nature Phys. {\bf 12}, 451 (2016).
\bibitem{Javanainen:96} J. Javanainen and S. M. Yoo, Phys. Rev. Lett. {\bf 76}, 161 (1996).
\bibitem{Castin:97} Y. Castin and J. Dalibard, Phys. Rev. A {\bf 55}, 4330 (1997).
\bibitem{Dziarmaga:03} J. Dziarmaga, Z. P. Karkuszewski and K. Sacha, J. Phys. B {\bf 36}, 1217 (2003).
\bibitem{Dagnino:09} D. Dagnino, N. Barber\'an and M. Lewenstein, Phys. Rev. A {\bf 80}, 053611 (2009).
 
 




%DD interaction







\end{thebibliography}
\end{document}